# Solar cycle changes of large-scale solar wind structure

P.K. Manoharan

Radio Astronomy Centre, National Centre for Radio Astrophysics,
Tata Institute of Fundamental Research, Udhagamandalam (Ooty), 643001, India
email: mano@ncra.tifr.res.in

**Abstract.** In this paper, I present the results on large-scale evolution of density turbulence of solar wind in the inner heliosphere during 1985–2009. At a given distance from the Sun, the density turbulence is maximum around the maximum phase of the solar cycle and it reduces to ∼70%, near the minimum phase. However, in the current minimum of solar activity, the level of turbulence has gradually decreased, starting from the year 2005, to the present level of ∼30%. These results suggest that the source of solar wind changes globally, with the important implication that the supply of mass and energy from the Sun to the interplanetary space has significantly reduced in the present low level of activity.

**Keywords.** turbulence, scattering, Sun: corona, Sun: evolution Sun: solar wind

## 1. Introduction

The solar wind plasma and its magnetic field undergo significant evolution with the phase of the solar cycle. In this paper, I present new results on the changing of large-scale structure of density turbulence in the solar wind over the solar cycles 22 and 23. The solar wind estimates have been obtained from the interplanetary scintillation (IPS) observations made with the Ooty Radio Telescope, operating at 327 MHz (e.g., Swarup et al. 1971). The IPS technique exploits the scattering of radio waves from a compact radio source by the density fluctuations in the solar wind (e.g., Hewish et al. 1964) and allows to probe the density turbulence in the solar wind, at a range of distances from the Sun as well as over an extended period of time.

## 2. Scintillation Index and Solar Wind Density Turbulence

The degree of interplanetary scintillation is given by the scintillation index, *m = rms of intensity fluctuation/mean intensity of the source*, which is a measure of density turbulence ($C_N^2(R) \sim m^2(R)$) in the solar wind. IPS measurements at Ooty can probe the solar wind in the heliocentric distance range of $R = 10 - 250$ solar radii ($R_\odot$) (e.g., Manoharan et al. 2000). Figure 1 shows the scintillation index variations with distance from the Sun, for the compact radio quasar 1148-001 between 1985 and 2008. In these plots, the enhanced scintillation caused by intense transient events (e.g., coronal mass ejections, refer to Manoharan et al. 2000) have been excluded and plots represent the average ambient turbulence condition of the heliosphere at different phases of the solar cycle.

The level of scintillation, as shown in the figure, increases to a peak value at a distance of $R \approx 40\ R_\odot$, and then decreases for further closer solar offsets, where the scattering becomes strong and saturated (e.g., Manoharan 1993). In the weak-scattering region at $R > 40\ R_\odot$, the gradual decrease in the level of scintillation is linearly related to the fall of density fluctuations ($C_N^2(R) \sim \delta N_e^2(R)$) with distance from the Sun and the radial dependence of scintillation index, $m = m_0 R^{-\alpha}$, can be related to the density turbulence, where $m_0$ corresponds to scintillation at unit distance from the Sun. However, an IPS measurement represents the integration along the line-of-sight to the radio source and therefore, the '*distance-density turbulence*' relationship, $C_N^2(R) \sim R^{-\beta}$, is related to the scintillation index by $2\alpha = (\beta - 1)$. In the above plots, the radial dependence of density turbulence can be obtained from the linear slope of the curve at distances >40 $R_\odot$ and slopes vary in the range between $\alpha \approx 2.0$ at the minimum and $\alpha \approx 1.5$ at the maximum of the cycle. These results indicate that the density turbulence falls off rapidly with distance, when the solar activity is low. Whereas it follows a typical $1/R^2$ dependence around the peak of the solar cycle. These results are in agreement with the earlier findings for solar cycles 21 and 22 (Manoharan 1993).

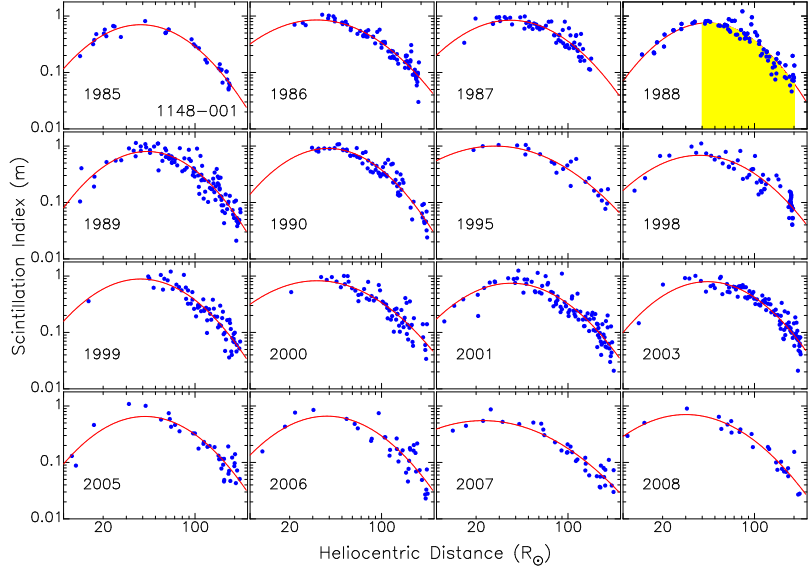

**Figure 1.** Scintillation index as a function of distance from the Sun for the radio quasar 1148-001 at 327 MHz, over the years 1985–2008. The peak of $m - R$ curve, close to unity, shows that the radio quasar 1148-001 is compact, which in fact has an equivalent diameter of $\sim$15 milli arcsec. In each plot, the best-fit to the data points is shown by a continuous curve.

### 2.1. *Solar Cycle Changes of Large-Scale Turbulence*

During the current minimum of activity, after about the year 2005, the peak of the scintillation curve however seems to move close to the Sun. It suggests that a given level of density turbulence moves closer to the Sun (i.e., a systematic reduction in turbulence at a particular distance from the Sun, from one year to the next). In order to determine the global characteristics of density turbulence, the area under the $m^2(R)$ profile has been calculated as given by,

$$A = \int_{R=40R_\odot}^{R=200R_\odot} m^2(R) \, dR. \quad (2.1)$$

The area has been estimated in the weak-scattering regime ($R \geqslant 40\ R_\odot$), which can provide a qualitative estimate of the large-scale turbulence in the inner heliosphere. Figure 2 shows the area of $m^2(R)$ curve plotted against the year for 10 compact sources. For each source, the area has been normalized by its fraction of scintillating flux density (i.e., the ratio of scintillating component flux and total flux density of the source), which is a constant value between unity and zero, respectively, corresponding to an ideal compact source and a broad non-scintillating source. It is evident that the integrated turbulence value peaks near the maximum of the solar cycle, around 1991 and a reduced value of ∼70% is observed at the minimum phase, around 1996. The interesting point is that at the maximum of cycle 23, turbulence level is less than that of the previous cycle maximum. Further, it decreases monotonously after the year 2005 to a level of ∼30% during 2008-09. A significant peak seen in density turbulence around the year 2003 is likely to be associated with the co-rotating interaction region dominated heliosphere, which has been caused by the persistent large mid-latitude coronal holes (Manoharan 2008).

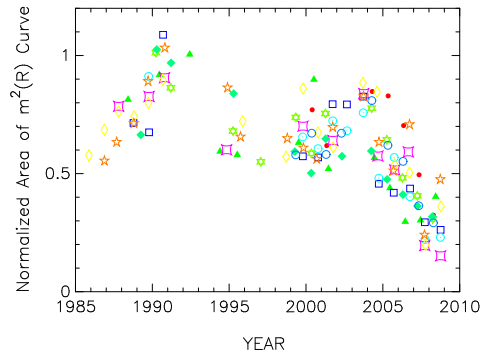

**Figure 2.** The area under the $m^2(R)$ profile (eqn. 2.1) plotted as a function of year. The area has be been calculated for the distance range of 40–200 $R_\odot$, as indicated by the shaded area in 1988 plot of Figure 1. Different symbols correspond to different scintillating sources.

## 3. Summary

The changes in the large-scale density turbulence obtained from the IPS measurements show that (1) the value of density turbulence is maximum at the maximum of the solar cycle; (2) it decreases to ∼70% level near the minimum; (3) however, during the current minimum of solar activity, the level of turbulence has gradually decreased to ∼30% of the previous cycle's maximum. The source of solar wind has changed globally and the reduction in density turbulence also correlates with the interplanetary magnetic field, solar wind speed and density. During the current minimum of solar activity, the heliosphere seems to face a reduction in the supply of mass and energy at the base of the corona.

### Acknowledgements

I acknowledge all the members of the Radio Astronomy Centre for the help in making IPS observations. This work is partially supported by the CAWSES-India Program, which is sponsored by ISRO.